\documentclass[12pt]{article}
\pdfoutput=1
\usepackage{bm}
\usepackage{mathrsfs}
\usepackage{amsbsy}
\usepackage{graphicx}
\usepackage{subfigure}
\usepackage{amsmath}
\usepackage{amssymb}
\usepackage[numbers,sort&compress]{natbib}

\newcommand{\mpref}[1]{Figure.\ref{#1}}
\numberwithin{equation}{section}
\begin{document}
	
	\begin{center}
		{\bf Island, Page Curve and Superradiance of Rotating BTZ Black Holes}\\
		
		\vspace{1.6cm}
		
		{\textbf{Ming-Hui Yu}$^{1}$,~\textbf{Cheng-Yuan Lu}$^{1}$, ~\textbf{Xian-Hui Ge}$^{1}$}\let\thefootnote\relax,
~\textbf{Sang-Jin Sin}$^{2}$\\
		\vspace{0.8cm}
		
		$^1${\it Department of Physics, Shanghai University, Shanghai 200444,  China} \\
        $^2${\it Department of Physics, Hanyang University, Seoul 04763, Korea}\\

		\vspace{1.6cm}

		\begin{abstract}
		We study the Page curve and the information paradox for the BTZ black hole coupled to two thermal baths by applying the island paradigm. We prove that as the island locates outside the event horizon, the entanglement entropy of Hawking radiation for finite temperature black holes obeys the Page curve. However, for extremal rotating black holes, the Page time and the scrambling time become divergent. To avoid the ill-defined of the Page time, we consider the contribution of superradiance, which is a process that can extract rotational energy from black holes. The superradiance continues in a period much shorter than the Page time when the central charge $c$ is not too large. In this process, the Page time, the scrambling time and the black hole thermal entropy all decrease. Whenever the superradiance finishes, the black hole turns out to be a neutral black hole and the Hawking radiation dominates. The Page curve can then be well reproduced.
		\end{abstract}
	\end{center}
\newpage
\tableofcontents
\newpage

\section{Introduction} \label{Introduction}
\qquad In 1975, Hawking proved that black holes can emit particles which are called Hawking radiation \cite{HR}. In this way, black holes can evaporate and disappear completely. However, the results show that Hawking radiation is a complete thermal spectrum. Suppose that a black hole is formed by the collapse of a pure state, and emits thermal Hawking radiation. Then it leads to the famous black hole information paradox \cite{Paradox}, because according to the unitarity principle, the final state should be a pure state rather than a mixed (thermal) state. This paradox can also be understood from the perspective of entropy. On the one hand, according to Hawking's calculations, the entanglement (fine-grained or von-Neumann) entropy of Hawking radiation in the process of black hole evaporation grows linearly and never stops. It also violates the finiteness of the von-Neumann entropy of a finite system. On the other hand, if the black hole evaporation is a unitary process, the entropy of radiation grows to the Bekenstein-Hawking (coarse-grained or thermodynamic) entropy \cite{entropy bound}, and then drops to zero when the black hole evaporates completely. Such a graph of the entanglement entropy vs time during the black hole evaporation is called the Page curve \cite{PC1,PC2}.\\
\indent Therefore, a key in solving the black hole information paradox is to reproduce the Page curve. Recently, there was a major breakthrough in this issue. Physicists obtained the Page curve through the ``island paradigm" for the first time in the framework of semi-classical physics \cite{island rule,entanglement wedge,review,bulk entropy}. Accordingly, the fine-grained entropy of Hawking radiation is given by the ``island formula" \citep{island rule}
\begin{subequations}
\begin{align}
S_{\text{Rad}}&=\text{Min} \{ \text{Ext} [S_{\text{gen}}] \},  \label{island formula} \\
 S_{\text{gen}}&=\frac{\text{Area}(\partial I)}{4G_N}+S_{\text{field}}(R \cup I).  \label{generalized entropy}
\end{align}
\end{subequations}
The second formula is called the generalized entropy, which is the extension of the Bekenstein-Hawking entropy \cite{Bekenstein entropy}, where $G_N$ represents the Newton constant, $R$ and $I$ stand for the radiation region and the island respectively. The boundary of the island is denoted as $\partial I$, i.e. the quantum extremal surface \cite{QES}. To obtain the first formula, we should first extremize the generalized entropy and find extremal points, which correspond to locations of the island. Then we choose the minimum value as the fine-grained entropy of radiation. Moreover, the island formula \eqref{island formula} can also be equivalently derived by the gravitational path integral \cite{replica1,replica2}. \\
\indent The Page curve was originally reproduced in two-dimensional evaporating black holes \cite{bulk entropy,island rule}. Meanwhile, the information paradox is also an important problem of eternal black holes. Compared with evaporating black holes, the information paradox of eternal black holes is more sharpened, and the requirements for the solution are more demanding \cite{eternal bh}. Therefore, we can understand the island paradigm more clearly by eternal black holes. Later, the island formula \eqref{island formula} was extensively applied to other backgrounds, such as Jackiw-Teitelboim (JT) gravity \cite{JT1,JT2,JT3,JT4,JT5}, two-dimensional asymptotically flat spacetime \cite{2d1,2d2,2d3,2d4,2d5}, higher-dimensional spacetime \cite{high1,high2,high3,high4,high5,high6,high7,high8,Sch,HV branes,flat space,RN,btz,dilaton,extremal,kk,charged dilaton1,charged dilaton2}, and gravitating baths \cite{bath1,bath2,bath3}. Furthermore, the island is also studied in the BCFT model \cite{BCFT1,BCFT2,BCFT3,BCFT4,BCFT5} and cosmology \citep{ds1,ds2,ds3,ds4,ds5}.\\
\indent In this paper, we study the Page curve for rotating Ba$\tilde{\text{n}}$ados-Teiteboim-Zanelli (BTZ) black holes which are coupled to two auxiliary flat bath systems. We emphasize the effect of angular momentum since it can make the black hole extremal. For non-extremal BTZ black holes, the Page curve can be reproduced. As the BTZ black hole approaches extremal, one will have trouble in reproducing the Page curve because there will be no Hawking radiation at zero temperature. To avoid such a situation, one should keep in mind that the superradiance reduces the angular momentum of a black hole so that a rotating black hole first loses its angular momentum because of the superradiance. The superradiance will reduce the Page time and at the end of it, the static BTZ black hole is restored. Therefore, the extremal case can be avoided.\\
\indent This paper is organized as follows. In section \ref{Set up}, we briefly review the property of BTZ black holes. In section \ref{Island paradigm for solving the information paradox}, we calculate the entanglement entropy of the radiation in the constructions without and with the island. The result with the island satisfies the Page curve. In section \ref{Superradiance}, we consider the superradiance in rotating black holes, which reduces the Page time and scrambling time so that the extremal case does not occur. Therefore, the problem of ill-definedness in the extremal case can be avoided. Lastly, we give the discussion and the conclusion in section \ref{Conclusion}. We set $\hbar=c=k_B=1$ hereafter.

\section{The model} \label{Set up}
\begin{figure}[htb]
\centering
\includegraphics[scale=0.35]{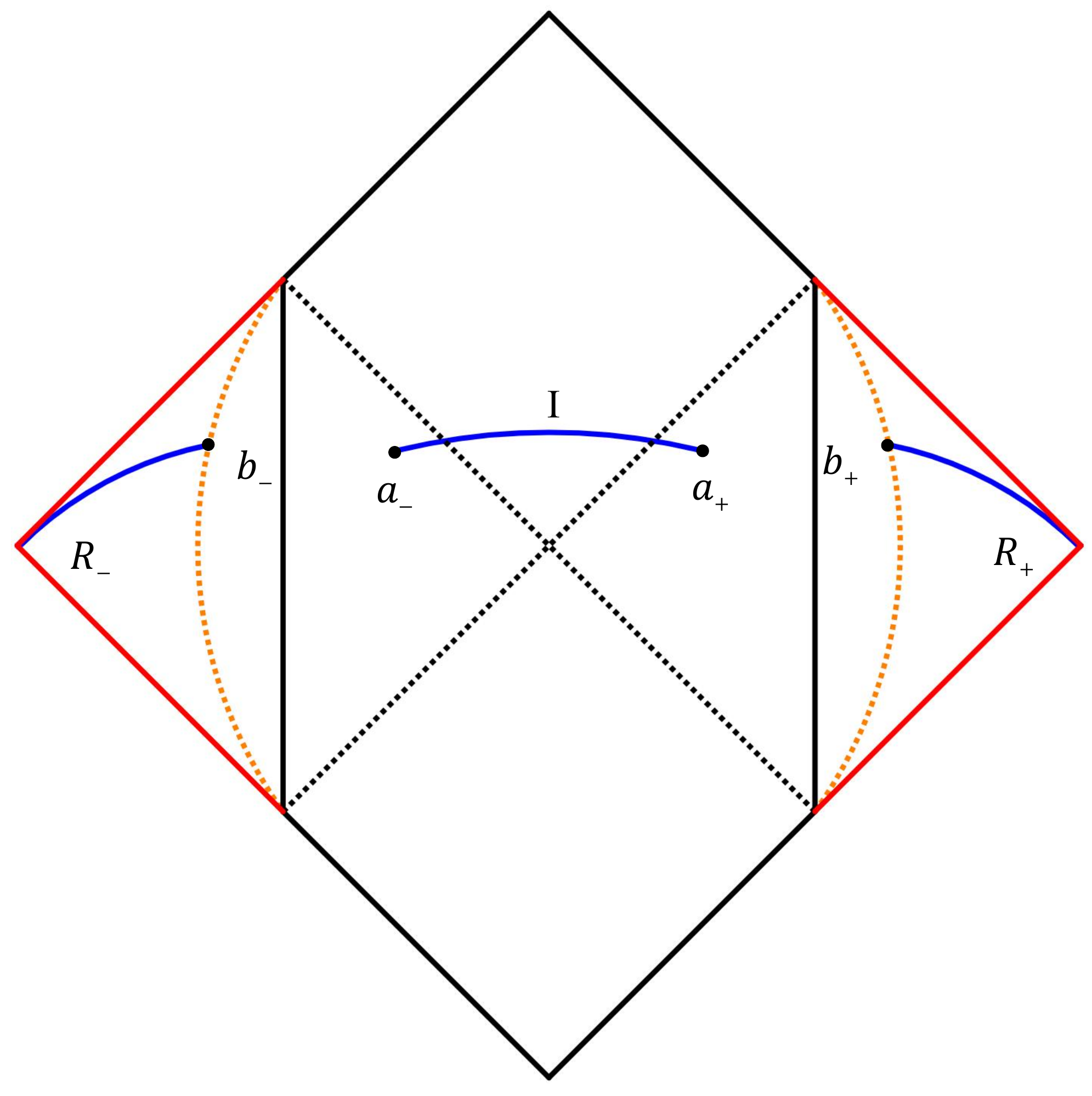}
\caption{\label{penrose1}Penrose diagram for non-extremal BTZ black hole in thermal equilibrium with baths. The black region represents the BTZ spacetime, and the red region represents the auxiliary flat spacetime. The orange dash lines refer to the cut-off surface. The entire system approximates an asymptotically flat black hole spacetime. The boundaries of radiation at cut-off surfaces are labeled as $b_{\pm}$, and the boundaries of the island are denoted as $a_{\pm}$.}
\end{figure}

\qquad In this section, we demonstrate how the two-sided rotating BTZ geometry is coupled to auxiliary flat spacetime (the bath system), i.e. the whole system is BTZ plus bath. The BTZ black hole is a special  solution of standard Einstein equations in (2+1)-dimensional spacetime \cite{BTZ bh}. Because there is no local degree of freedom in the three-dimensional spacetime and those with negative curvature are locally $\text{AdS}_3$. This is very different from the asymptotically flat Kerr spacetime. Nevertheless, it shares many features with (3+1)-dimensional black holes, and because the structure is simple enough, the involved calculations can be analytically performed (see \cite{btz review} for an exhaustive review of this black hole). Thus, it is natural to study the black hole information paradox in BTZ black holes.\\
\indent The action of the whole system is given by \cite{BTZ bh}
\begin{subequations}
\begin{align}
I&=I_{\text{grav}}+I_{\text{matter}},   \label{action1} \\
I_{\text{grav}}&=\frac{1}{16 \pi G_N} \int d^3 x \sqrt{-g} (R+2 \Lambda), \label{action2}
\end{align}
\end{subequations}
where $R$ is the three-dimensional Ricci scalar curvature which is a constant everywhere. The AdS length $\ell$ is related to the negative cosmological constant $\Lambda$ by $\Lambda \equiv -\frac{1}{\ell^2}$. The $I_{\text{matter}}$ is the action of CFT that lives in the boundary. The metric in the gravity region has the following form \cite{BTZ bh}
\begin{equation}
ds^2=-f(r)dt^2+f^{-1}(r)dr^2+r^2(d \phi +N^{\phi}dt)^2,  \label{metric1}
\end{equation}
where
\begin{equation}
f(r)=-8G_NM+\frac{r^2}{\ell^2}+\frac{16G_N^2J^2}{r^2}=\frac{(r^2-r_+^2)(r^2-r_-^2)}{\ell^2r^2}, \label{metric function}
\end{equation}
and
\begin{equation}
N^{\phi}=-\frac{4G_NJ}{r^2}=\frac{r_+r_-}{\ell r^2}. \label{angular function}
\end{equation}
Hereafter we set the AdS radius $\ell =1$ for convenience. $M$ and $J$ stand for the mass and the angular momentum of black holes. The two parameters $r_{\pm}$ represent the event horizon and the inner horizon of the black hole, which can be written as
\begin{equation}
r_{\pm} = \sqrt{2G_N (M+J)} \pm \sqrt{2G_N (M-J)}.  \label{horizons}
\end{equation}
In order to ensure the existence of the event horizon, we must have $|J| \le M$ and $M>0$. In the extremal case $|J|=M$, two horizons coincide. The Hawking temperature of the BTZ black hole is
\begin{equation}
T_H = \frac{1}{\beta} = \frac{r_+^2-r_-^2}{2 \pi r_+}. \label{temperature}
\end{equation}
The Bekenstein-Hawking entropy read as
\begin{equation}
S_{BH}= \frac{\pi r_+}{2G_N}. \label{bh entropy}
\end{equation}
\indent To define the spacelike periodic coordinate, we introduce the co-moving coordinate
\begin{equation}
\tilde{\phi} = \phi - \Omega_H t,   \label{comoving coordinate}
\end{equation}
where $\Omega_H$ is the angular velocity of the event horizon
\begin{equation}
\Omega_H = -N^{\phi} (r_+) = \frac{r_-}{r_+}.  \label{angular velocity}
\end{equation}
Then, the metric \eqref{metric1} of the two-dimensional surface with $\tilde{\phi} = \text{constant}$ can be rewritten in the following form
\begin{equation}
ds^2=-\frac{(r^2-r_+^2)(r^2-r_-^2)}{r^2}dt^2+\frac{r^2}{(r^2-r_+^2)(r^2-r_-^2)}dr^2.  \label{metric2}
\end{equation}
Besides, in order to obtain the maximum extension of the BTZ spacetime, one can define the Kruskal coordinate
\begin{subequations}
\begin{align}
\text{Right Wedge}: \ \  &U=-e^{-\kappa (t-r^{\ast})}, \qquad V=+e^{+\kappa(t+r^{\ast})},     \label{Kruskal1} \\
\text{Left Wedge}: \ \   &U=+e^{-\kappa (t-r^{\ast})}, \qquad V=-e^{+\kappa(t+r^{\ast})},    \label{Kruskal2}
\end{align}
\end{subequations}
where $\kappa=\frac{2\pi}{\beta}$ is the surface gravity, and $r^{\ast}$ denotes the tortoise coordinate
\begin{equation}
r^{\ast}(r)= \int f^{-1}(r)dr =\frac{1}{2(r_+^2-r_-^2)} \Bigg[ r_+ \log \frac{|r-r_+|}{r+r_+} -r_- \log \frac{|r-r_-|}{r+r_-} \Bigg]. \label{tortoise}
\end{equation}
\indent Now, we make the boundary of BTZ spacetime transparent by using the technique of coupling auxiliary baths \cite{auxiliary bath}. For the bath, we can regard it as a flat Minkowski spacetime without the gravitational effect and assume it is in thermal equilibrium with the black hole. The corresponding Penrose diagram is shown in \mpref{penrose1}.\\
\indent For the whole spacetime (BTZ + bath), its metric can be written in terms of the Kruskal coordinate as follows
\begin{equation}
ds^2=-\Omega^2dUdV,   \label{mapping}
\end{equation}
where
\begin{equation}
\Omega_{\text{BTZ}}=\frac{\sqrt{f(r)}}{\kappa} e^{-\kappa r^{\ast}(r)}, \qquad \Omega_{\text{bath}}=\frac{1}{\kappa}e^{-\kappa r},  \label{conformal factor}
\end{equation}
are conformal factors. Then, the geodesic distance $d(x_1,x_2)$ between two points in the whole spacetime can be written as follows
\begin{equation}
d(x_1,x_2)=\Omega(x_1) \Omega (x_2) \bigg[ U(x_2) -U(x_1) \bigg] \bigg[ V(x_1)-V(x_2) \bigg].  \label{geodesic}
\end{equation}
Before specific calculations, we emphasize two calculational tricks. \\
(i) The entanglement entropy in higher-dimensional spacetime has area-like UV divergent, which depends on the cut-off $\epsilon$ and can be absorbed by renormalization of the Newton constant\footnote{For the convenience of future writing, we omit the superscripts $(\text{fin})$ and $(r)$.} $G_N^{(r)}$
\begin{subequations}
\begin{align}
S_{\text{field}} (R \cup I) &=\frac{\text{Area}(\partial I)}{\epsilon} +S_{\text{field}}^{(\text{fin})} (R \cup I),  \\
\frac{1}{4G_N^{(r)}} &= \frac{1}{4G_N}+\frac{1}{\epsilon}.
\end{align}
\end{subequations}
(ii) Through the angular symmetry analysis in the direction of $\tilde{\phi}$ \eqref{comoving coordinate}, we can calculate the entanglement entropy of the QFT in disconnected intervals. There exist a reduced QFT of massless fermions in the two-dimensional geometry\footnote{In fact, it also includes the massive Kaluze-Klein mode, but we ignore its contribution to the entanglement entropy at the long distance.} \eqref{metric2} \cite{EE formula}
\begin{equation}
S_{\text{field}} (R \cup I)= \frac{c}{6} \log \Bigg[ \frac{d(a_+,a_-)d(b_+,b_-)d(a_+,b_+)d(a_-,b_-)}{d(a_+,b_-)d(a_-,b_+)}   \Bigg],   \label{EE formula}
\end{equation}
where $d(x,y)$ is defined in \eqref{geodesic}. Here we set the central $c$ is not very large: $c \ll \frac{1}{G_N}$.

\section{Island and Page time for rotating BTZ black holes} \label{Island paradigm for solving the information paradox}
\qquad In this section, we calculate the entanglement entropy of Hawking radiation by the island formula \eqref{island formula} and reproduce the Page curve of BTZ black holes to solve the information paradox. We only consider the non-extremal case in this section.\\
\indent Being different from the asymptotically flat spacetime, the reflecting boundary conditions of AdS prevent large black holes in the asymptotically AdS spacetime from evaporating completely. In order to simulate an evaporation process for black holes in the asymptotically  AdS spacetime, we couple baths to the BTZ spacetime (\mpref{penrose1}) and let them in thermal equilibrium. Then the whole system is a Hartle-Hawking state at $t=0$ and evolves in time. Now, the Hawking radiation can be collected in bath regions and allows us to evaluate its fine-grained entropy by island formula \eqref{island formula}.

\subsection{Absence of Island at early times}
\qquad In this subsection, we prove that the island does not appear at early times, which causes the entanglement entropy to increases linearly in time. Assuming that the system is a pure state at $t=0$. The endpoints of boundaries of radiation regions $R_{\pm}$ and the island region $I$ are denoted as $b_{\pm}=(\pm t_b,b)$ and $a_{\pm} =(\pm t_a,a)$ respectively. Accordingly, the contribution of quantum matter fields is obtained by \eqref{geodesic} and \eqref{EE formula}
\begin{equation}
\begin{split}
S_{\text{field}}(R \cup I) &=\frac{c}{6} \log \Bigg[ 16e^{2\kappa(r^{\ast}(a)+b)} \cosh^2(\kappa t_a) \cosh^2(\kappa t_b) \Omega^2_{\text{BTZ}}(a) \Omega^2_{\text{bath}}(b) \Bigg]\\
&+\frac{c}{3} \log \Bigg[\frac{\cosh[\kappa(r^{\ast}(a)-b)]-\cosh[\kappa (t_a-t_b)]}{\cosh[\kappa(r^{\ast}(a)-b)]+\cosh[\kappa (t_a-t_b)]} \Bigg]. \label{field EE}
\end{split}
\end{equation}
At early times, $t_a,t_b \ll b$, then the above expression simplifies to
\begin{equation}
S_{\text{field}}^{(\text{early})} \simeq \frac{c}{6} \log \Bigg[ \frac{16f(a)}{\kappa^4} \cosh^2 (\kappa t_a) \cosh^2 (\kappa t_b) \Bigg]. \label{early field EE}
\end{equation}
so the generalized entropy at early times read as
\begin{equation}
S_{\text{gen}}^{(\text{early})}= \frac{\pi a}{G_N} +\frac{c}{6} \log \Bigg[ \frac{16f(a)}{\kappa^4} \cosh^2 (\kappa t_a) \cosh^2 (\kappa t_b) \Bigg]. \label{early gen EE}
\end{equation}
We can obtain the location of the island by extremizing the above equation. The extremization with respect to $a$ and $t_a$ is follow
\begin{equation}
\begin{split}
\frac{\partial S_{\text{gen}}^{(\text{early})}}{\partial a}&= \frac{\pi}{G_N} +\frac{c}{6} \frac{f^{\prime}(a)}{f(a)}, \\
&= \frac{\pi}{G_N} + \frac{c}{3} \frac{a^4-r_+^2r_-^2}{a(a^2-r_+^2)(a^2-r_-^2)}=0,  \label{early gen EE a}
\end{split}
\end{equation}
and
\begin{equation}
\frac{\partial S_{\text{gen}}^{(\text{early})}}{\partial t_a}=\frac{c}{3} \frac{\kappa \sinh(\kappa t_a)}{\cosh(\kappa t_a)}=0. \label{early gen EE t}
\end{equation}
Then the boundary of the island is
\begin{equation}
a \simeq \frac{c}{3\pi} \cdot \ell_p, \qquad t_a=0,  \label{early island location}
\end{equation}
where $\ell_p \equiv \sqrt{G_N}$ is the Planck length. We should discard all Planck scale physics, which implies that there is no island at early times. Therefore, in the absence of the island, only the contribution of quantum fields to the entanglement entropy is left in the generalized entropy \eqref{generalized entropy}, which is written as\footnote{It can be obtained simply from the formula \eqref{EE formula}. More precisely, according to the complementarity of the von Neumann entropy, the entanglement entropy of the radiation region is equal to the region $[b_-,b_+]$ in the construction without island.}
\begin{equation}
S_{\text{gen}}(\text{without \ island }) \equiv S_{\text{field}}(R) =\frac{c}{6} \log [d(b_+,b_-)] =\frac{c}{3} \log \bigg[\frac{2}{\kappa} \cosh (\kappa t_b) \bigg]. \label{EE without island}
\end{equation}
However, at late times
\begin{equation}
\cosh(\kappa t_b) \simeq \frac{1}{2} e^{\kappa t_b}. \label{approximate1}
\end{equation}
Then
\begin{equation}
S_{\text{gen}} (\text{without \ island}) \simeq \frac{c}{3} \kappa t_b.  \label{result1}
\end{equation}
\indent The result of this linear growth leads to the information paradox for the BTZ black hole since the black hole has the only finite degree of freedom, which is at most the Bekenstein-Hawking entropy \cite{entropy bound}. At late times, black holes do not have enough degrees of freedom to purify Hawking radiation. In order to resolve this paradox, Page indicates that one has to take account of the outgoing Hawking quanta and the interior partner together since they are entangled with each other, forming a pure state. In the language of the entanglement wedge reconstruction \cite{entanglement wedge}, we need to introduce a new region--``island" in the entanglement wedge of the radiation. Therefore, we expect that the island paradigm can solve this paradox and give the Page curve at late times.

\subsection{Island outside the event horizon at late times}
\qquad In this subsection, we show that the island emerges at late times, which can stop the growth of radiation entropy. The goal is to find the location of the island. At late times $t_a,t_b \gg \kappa$, and the distance between the left wedge and the right in \mpref{penrose1} is very large, so we have
\begin{equation}
d(a_+,a_-) \simeq d(b_+,b_-) \simeq d(a_{\pm},b_{\mp}) \gg d(a_{\pm},b_{\pm}). \label{OPE limit}
\end{equation}
the expression \eqref{EE formula} reduces as
\begin{equation}
S_{\text{field}}^{(\text{late})} \simeq \frac{c}{3} \log \Bigg[ 2e^{\kappa (r^{\ast}(a)+b} \Omega_{\text{BTZ}}(a) \Omega_{\text{bath}}(b) \bigg( \cosh[\kappa (r^{\ast}(a)-b)] -\cosh[\kappa (t_a-t_b)] \bigg)    \Bigg].  \label{late field EE}
\end{equation}
So the corresponding generalized entropy is
\begin{equation}
S_{\text{gen}}^{(\text{late})}= \frac{\pi a}{G_N}+\frac{c}{3} \log \Bigg[ 2e^{\kappa (r^{\ast}(a)+b} \Omega_{\text{BTZ}}(a) \Omega_{\text{bath}}(b) \bigg( \cosh[\kappa (r^{\ast}(a)-b)] -\cosh[\kappa (t_a-t_b)] \bigg)    \Bigg]. \label{late gen EE}
\end{equation}
By extremizing the above expression with respect to $t_a$ at first
\begin{equation}
\frac{\partial S_{\text{gen}}^{(\text{late})}}{\partial t_a} =\frac{c}{3} \frac{- \kappa \sinh[\kappa (t_a-t_b)]}{\cosh[\kappa (r^{\ast}(a)-b)]-\cosh[\kappa(t_a-t_b)]}=0. \label{late gen EE t}
\end{equation}
The only solution for the above equation is $t_a=t_b$, thus the extremization with respect to $a$ by invoking $t_a=t_b=t$, we obtain
\begin{equation}
\frac{\partial S_{\text{gen}}^{(\text{late})}}{\partial a}= \frac{c}{6} \frac{f^{\prime}(a)(\cosh[\kappa (r^{\ast}(a)-b)]-1)+2\kappa \sinh [\kappa (r^{\ast}(a)-b)]}{f(a)(\cosh[\kappa (r^{\ast}(a)-b)]-1)}. \label{late gen EE a}
\end{equation}
By setting $r \sim r_+$, the quantum extremal surface is at
\begin{equation}
\begin{split}
a &\simeq r_+ + \frac{c^2G_N^2}{18 \pi^2 r_+} \bigg( \frac{r_+-r_-}{2r_+} \bigg)^{-\frac{r_-}{r_+}}e^{2 \kappa b}, \\
&=r_+ + {\cal O} \bigg( \frac{c^2G^2_N}{r_+}\bigg).  \label{island location}
\end{split}
\end{equation}
Finally, the fine-grained entropy of Hawking radiation at late times is given by
\begin{equation}
\begin{split}
S_{\text{Rad}}&=\frac{\pi r_+}{G_N} +\frac{c}{6} \log \Bigg[ \frac{r_+^2}{(r_+^2-r_-^2)^3} \Bigg] + \cdots \\
&\simeq 2S_{BH}. \label{result2}
\end{split}
\end{equation}
The leading order is the Bekenstein-Hawking entropy, which comes from the contribution of the island. The subleading order and other terms are the quantum effects from matter fields, which can be omitted compared with the first term. Therefore, the entropy of radiation reaches an asymptotic constant, which implies the construction with the island is correct.\\
\indent For the process, without island at early times, the generalized entropy \eqref{generalized entropy} keeps increasing. But as soon as we consider the island outside the event horizon at late times, the entropy stops growing. Therefore, the entanglement entropy of Hawking radiation is consistent with the unitary Page curve.

\subsection{Page time and Scrambling time}
\qquad Now, we give the Page time and the scrambling time in this subsection. The Page time is defined as the time when the entanglement entropy of Hawking radiation reaches its peak. For an eternal black hole, the entropy does not change after it. By equating the entanglement entropy without island \eqref{result1} with the entropy with island at late times \eqref{result2}, the Page time is determined as
\begin{equation}
t_{\text{Page}} =\frac{6S_{BH}}{c \kappa} =\frac{3\beta}{\pi c} S_{BH}.  \label{page time}
\end{equation}
The shortest time during which the information can be recovered from the Hawking radiation is defined as the scrambling time by the Hayden-Preskill experiment \cite{HP test}. In the entanglement wedge reconstruction proposal \cite{entanglement wedge}, the scrambling time corresponds to the time for the information to enter the island, since the degree of freedom of the island belongs to radiation. Suppose an observer on the cut-off surface ($r=b$) sends a light signal at time $t_1$. It arrives at the island ($r=a$) at time $t_2$. Accordingly, the distance in the null direction is
\begin{equation}
V(t_1,b)-V(t_2,a)=(t_1+r^{\ast}(b))-(t_2+r^{\ast}(a)).
\end{equation}
and the delivery time is
\begin{equation}
t_2-t_1=[r^{\ast}(b)-r^{\ast}(a)]-[V(t_1,b)-V(t_2,a)].
\end{equation}
The shortest time for information recovery is
\begin{equation}
\begin{split}
t_{\text{scr}}\equiv r^{\ast}(b)-r^{\ast}(a) &\simeq \frac{1}{2(r_+^2-r_-^2)} \Bigg[ r_+ \log \Bigg[ \frac{(b-r_+)r_+}{c^2G_N^2} \Bigg] -r_- \log \bigg( \frac{b-r_-}{r_+-r_-} \bigg) \Bigg] \\
&\simeq \frac{r_+}{(r_+^2-r_-^2)} \log \bigg( \frac{r_+}{cG_N} \bigg) \simeq \frac{\beta}{2\pi} \log S_{BH}. \label{scrambling time}
\end{split}
\end{equation}
Here, we set $b$ to have the same order with $r_+$ and assume that $S_{BH} \simeq \frac{r_+}{G_N}$. This result is consistent with \cite{scrambling time}.

\section{Superradiance and Page time} \label{Superradiance}
\begin{figure}[htb]
\centering
\subfigure[\scriptsize{}]{\label{angualr}
\includegraphics[scale=0.27]{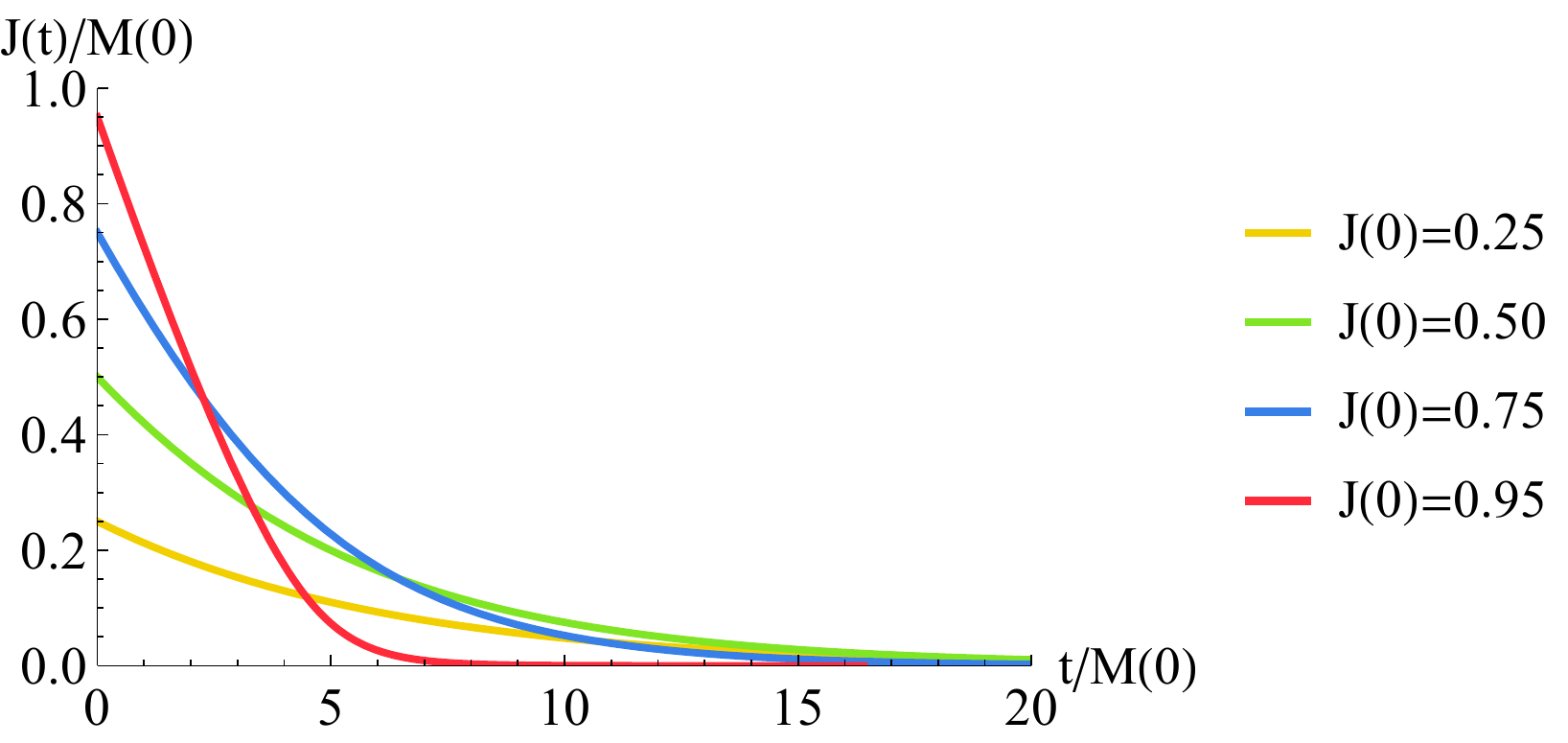}
}
\quad
\subfigure[\scriptsize{}]{\label{mass}
\includegraphics[scale=0.27]{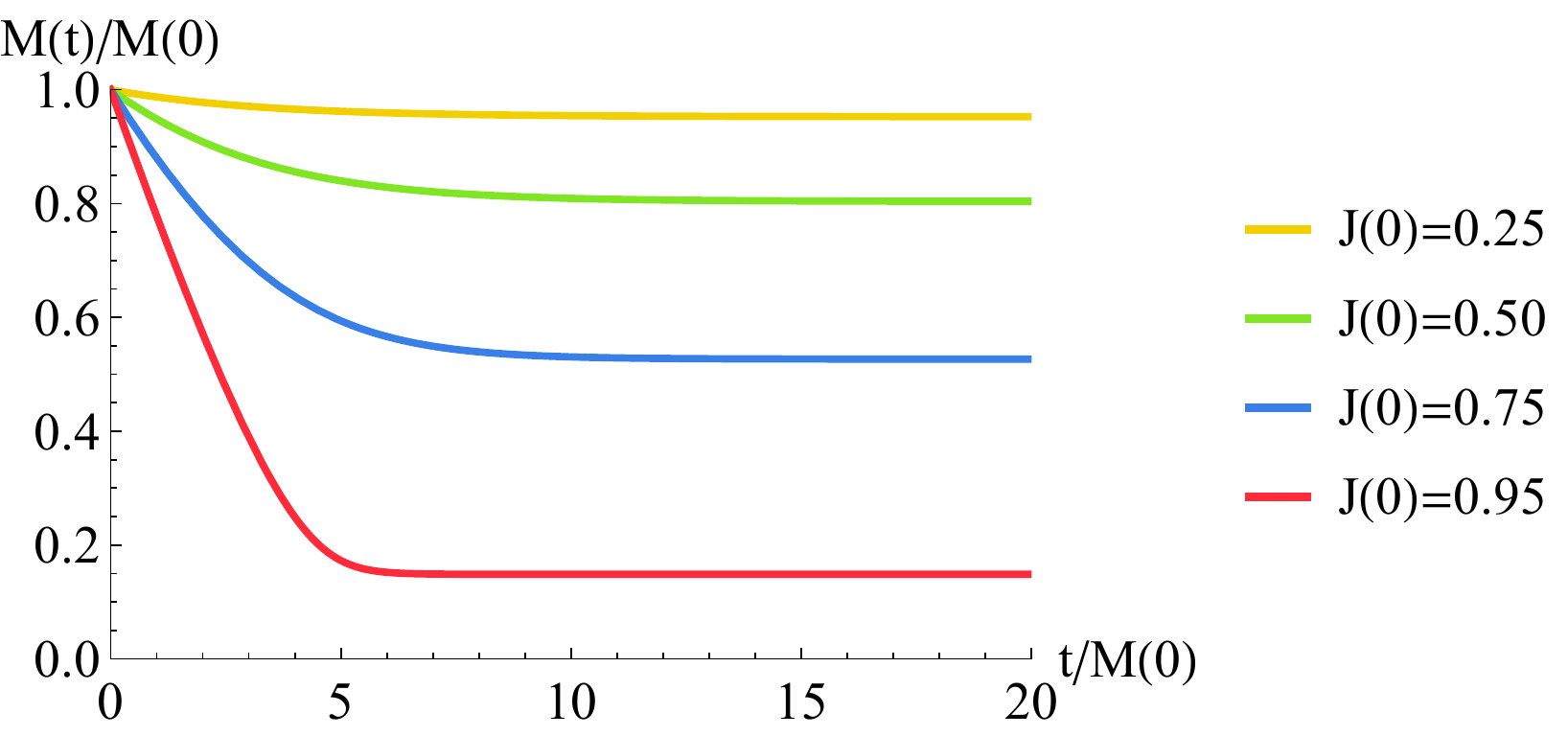}
}
\quad
\subfigure[\scriptsize{}]{\label{BH entropy}
\includegraphics[scale=0.27]{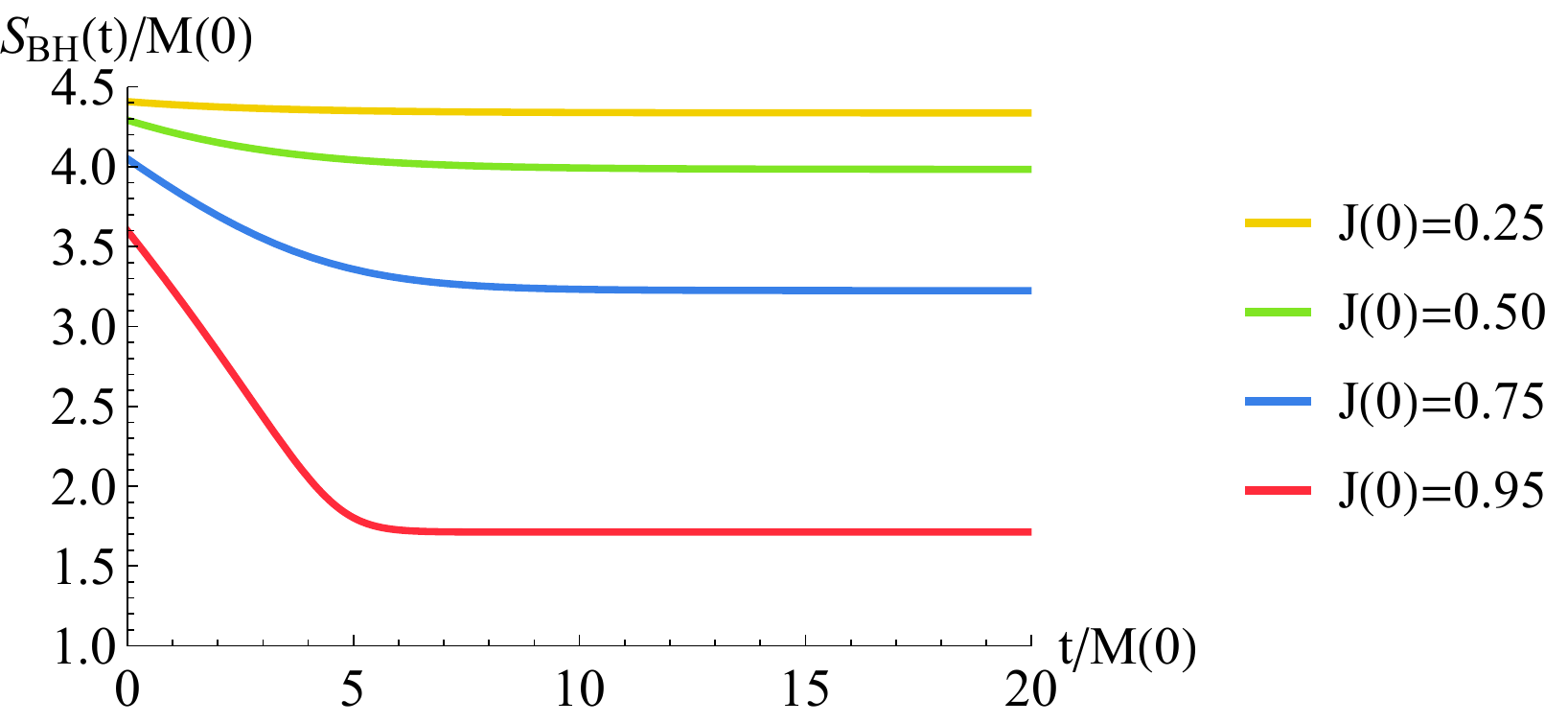}
}

\caption{The parameters of the system as a function for the time $t$ in the evaporation process. Here we set $G_N=1$.}
\label{bh parameters}
\end{figure}

\begin{figure}[htb]
\centering
\subfigure[\scriptsize{}]{\label{tpage}
\includegraphics[scale=0.57]{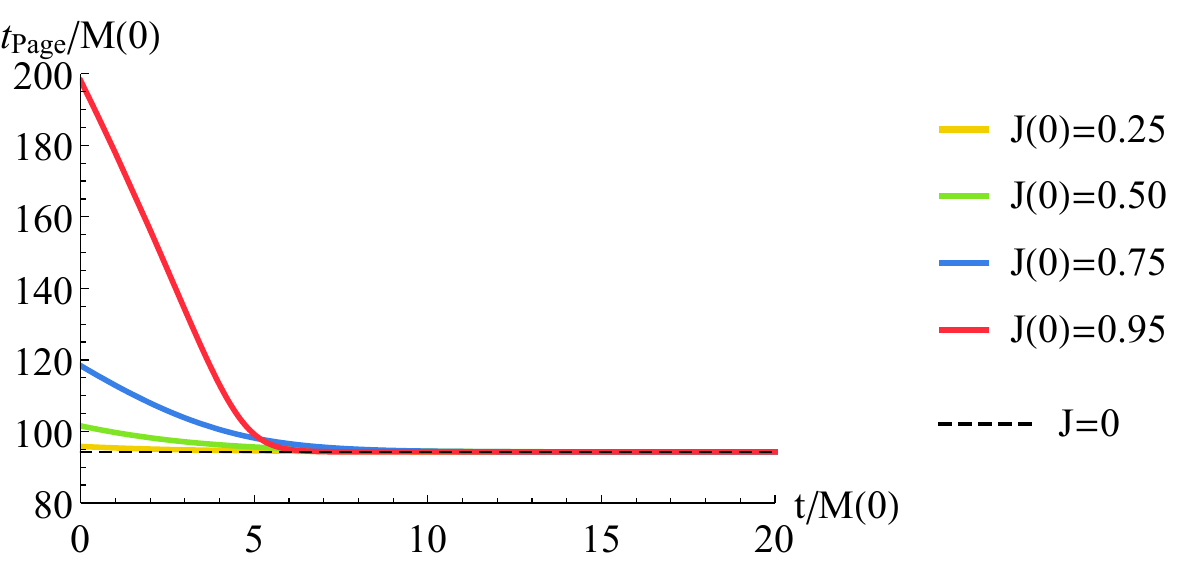}
}
\quad
\subfigure[\scriptsize{}]{\label{tscr}
\includegraphics[scale=0.4]{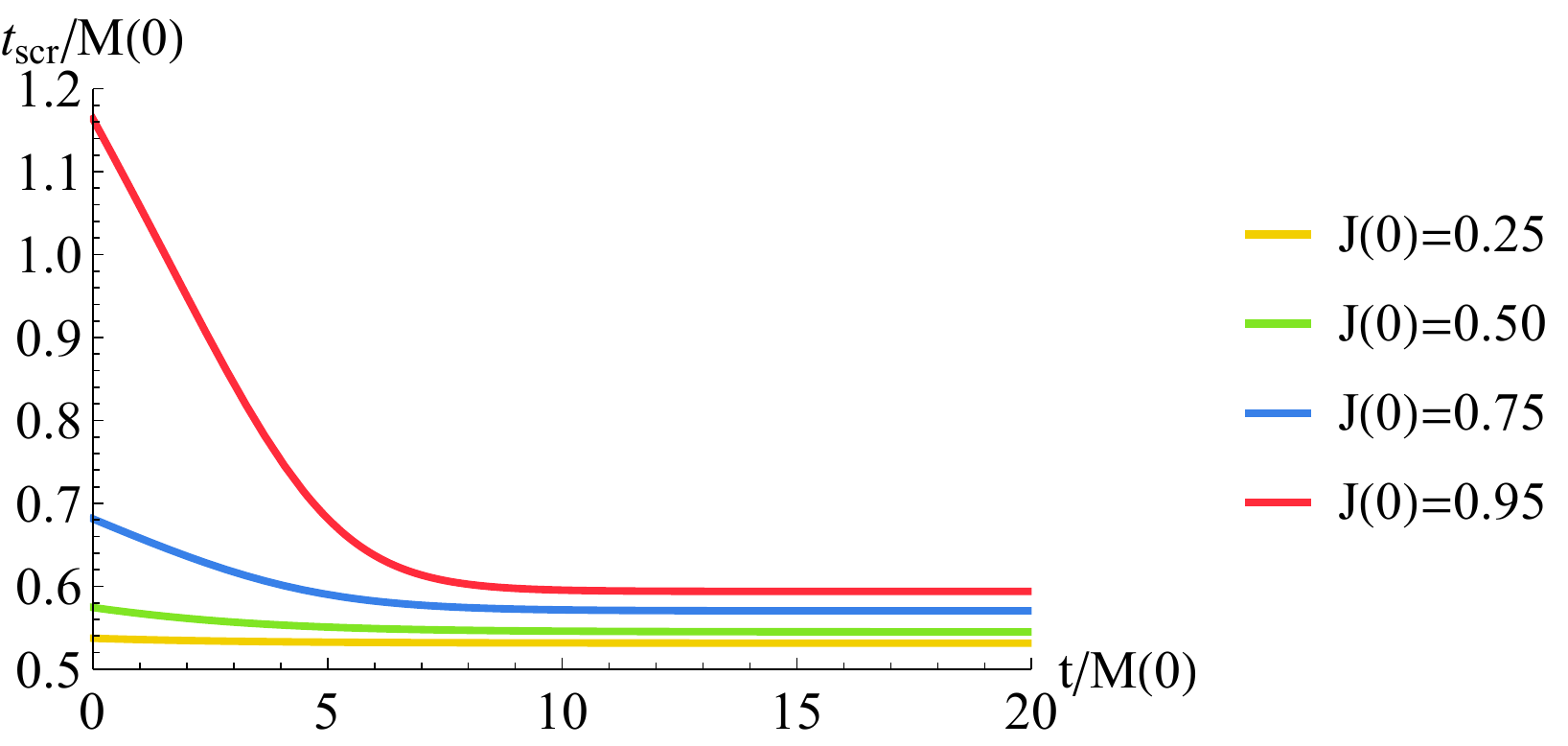}
}

\caption{The Page time and the scrambling time as a function for the $t$. Here we set the central charge $c=0.1 \ll 1/G_N=1$.}
\label{tp and tscr}
\end{figure}

\begin{figure}[htb]
\centering
\subfigure[\scriptsize{}]{\label{j1}
\includegraphics[scale=0.5]{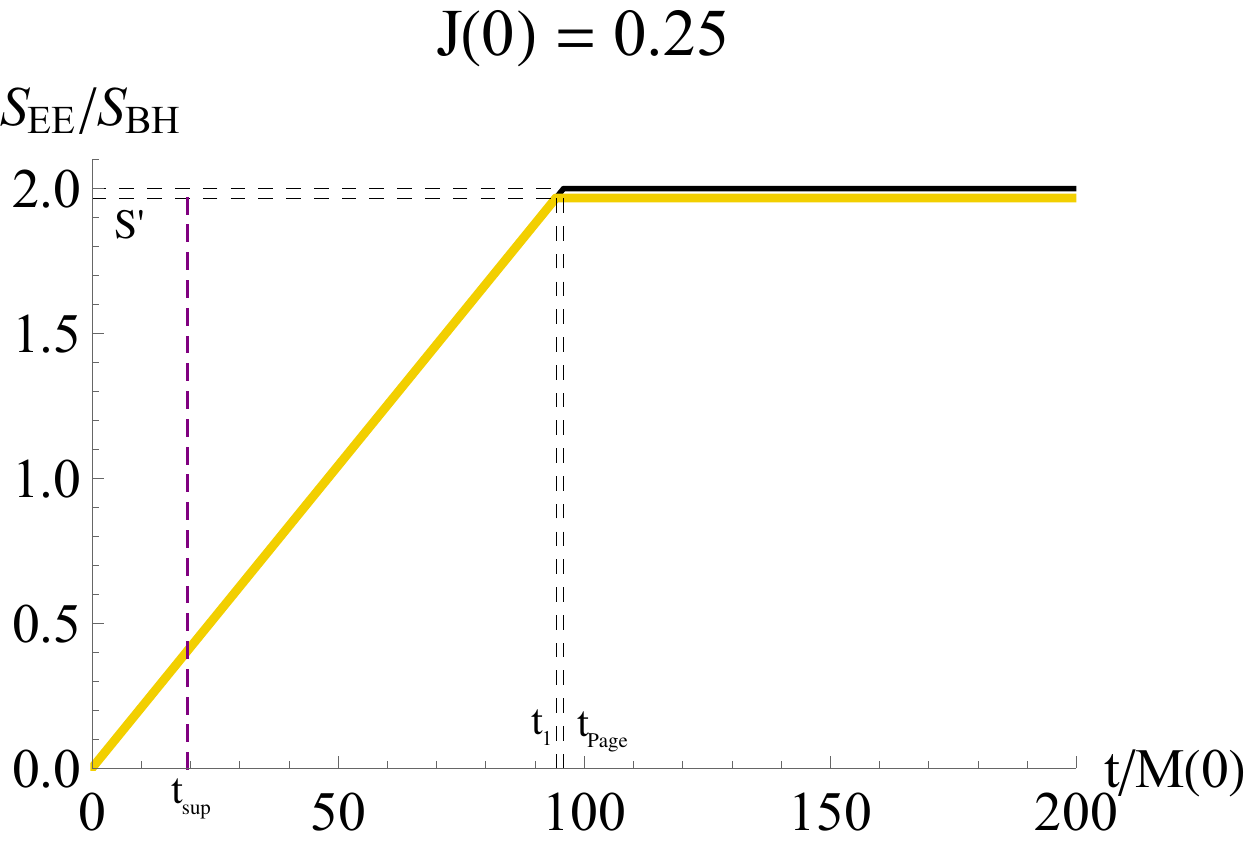}
}
\quad
\subfigure[\scriptsize{}]{\label{j2}
\includegraphics[scale=0.5]{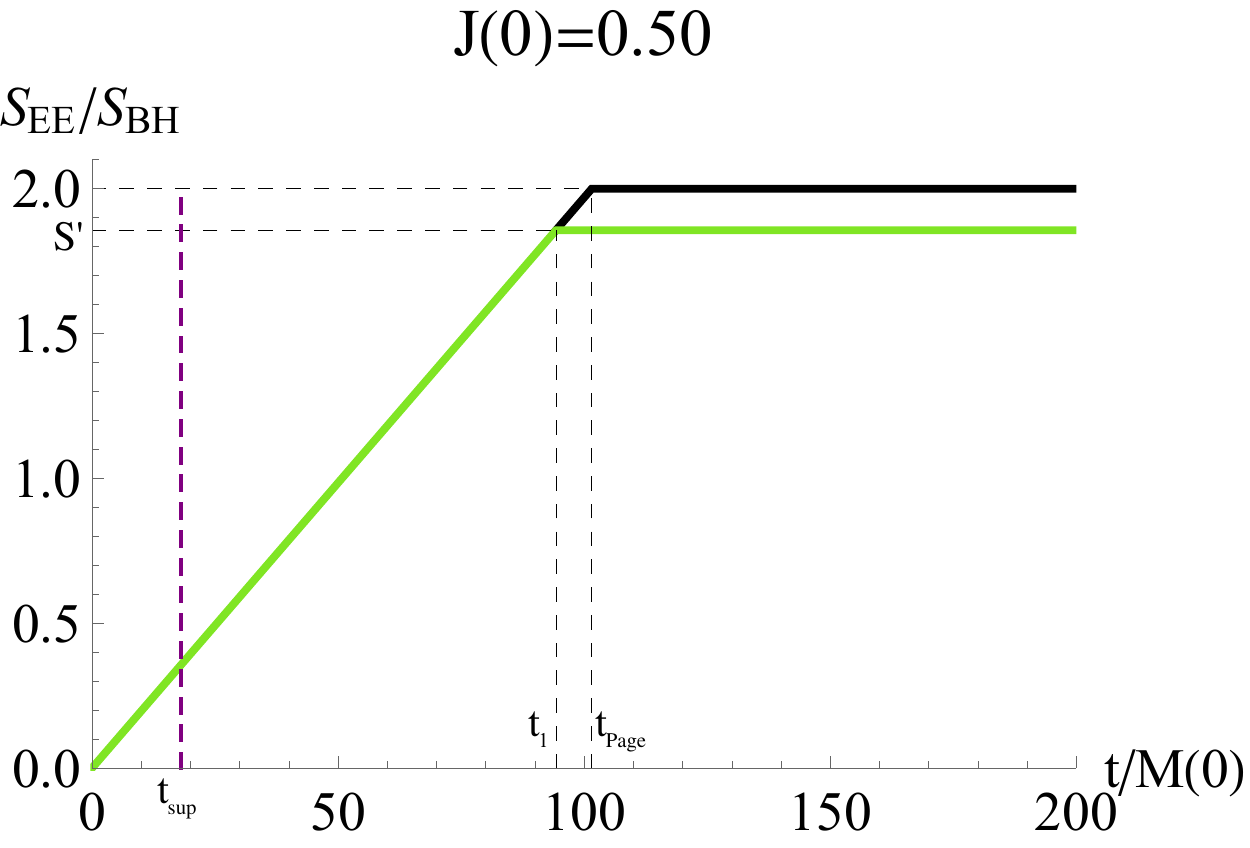}
}
\quad
\subfigure[\scriptsize{}]{\label{j3}
\includegraphics[scale=0.5]{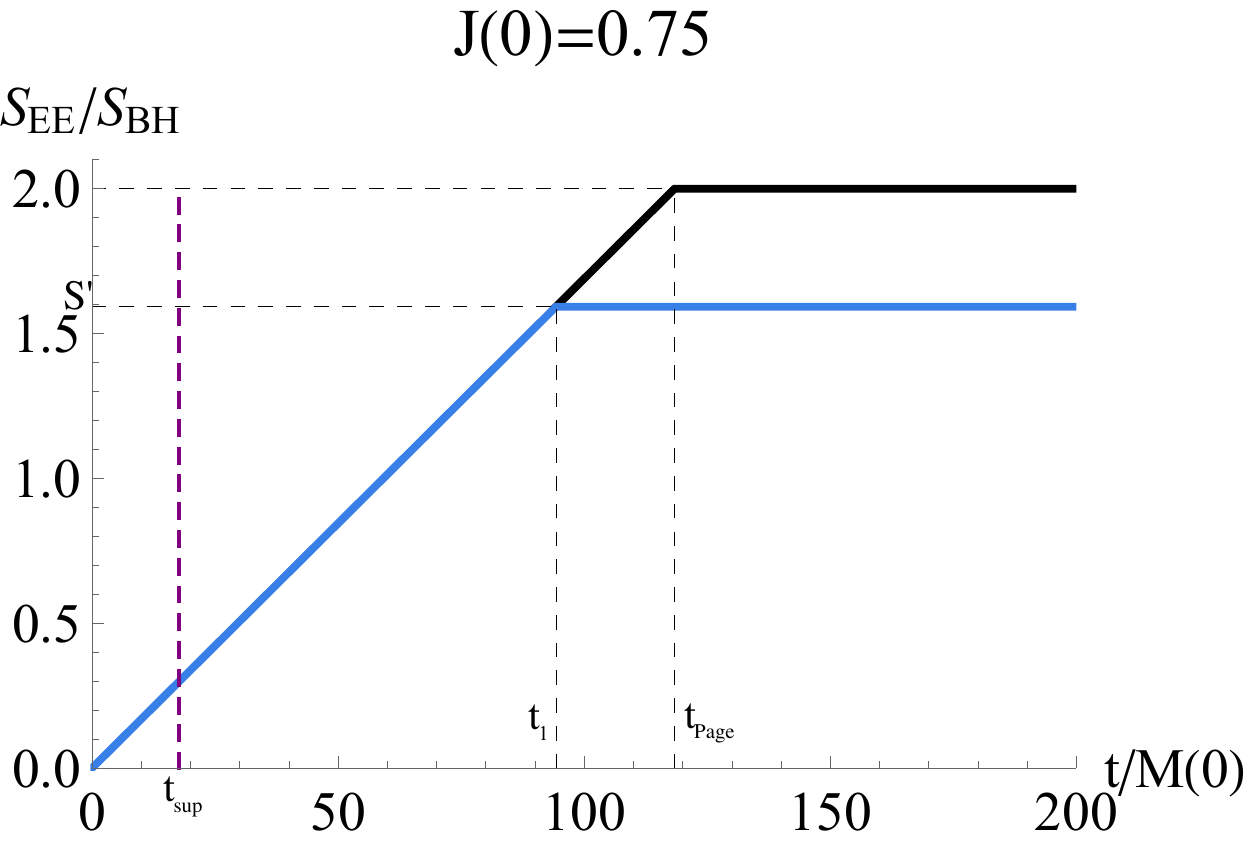}
}
\quad
\subfigure[\scriptsize{}]{\label{j4}
\includegraphics[scale=0.5]{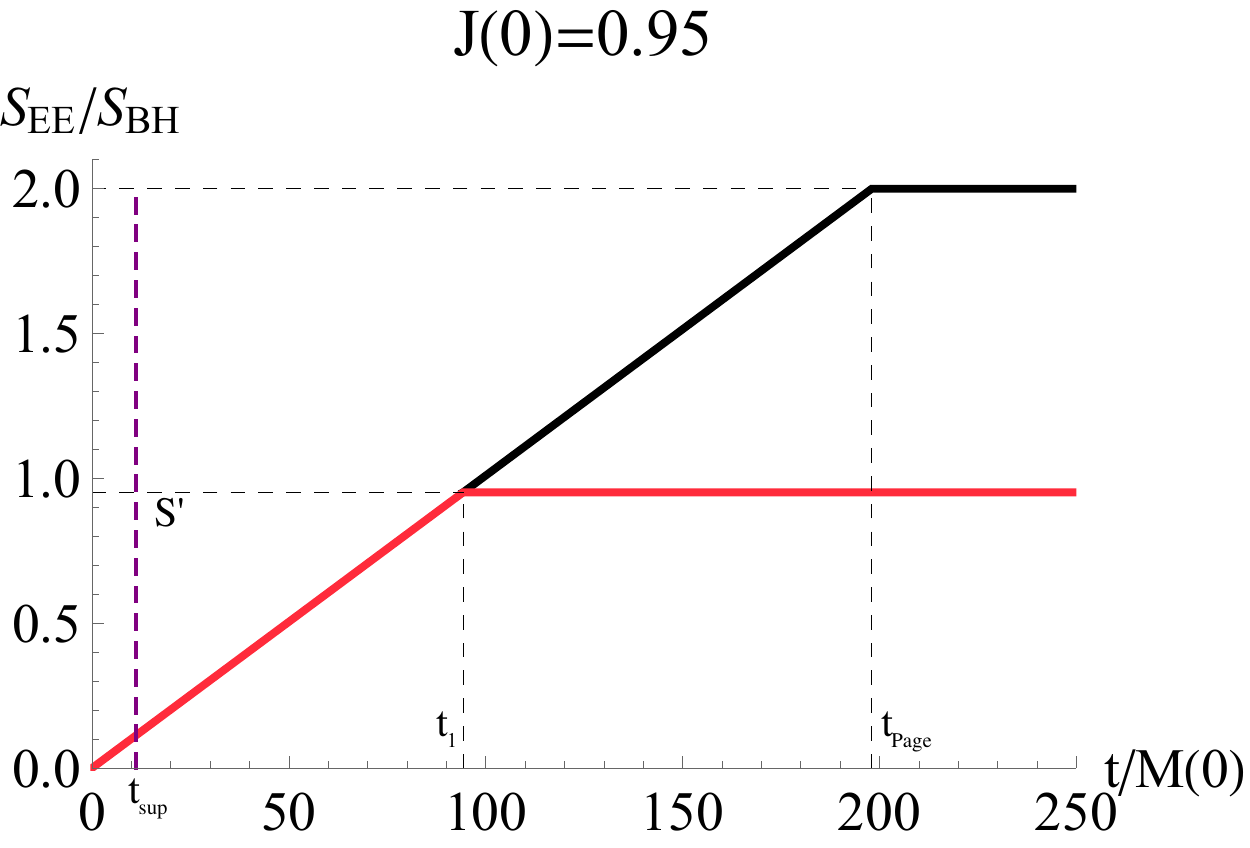}
}
\caption{The Page curve of the system. Black lines represents the situation where the superradiance is not considered. $S_{BH}$ and $t_{\text{Page}}$ are the original Page time and Bekenstein-Hawking entropy, and they becomes $S^{\prime}$ and $t_1$ after considering the superradiance. These purple lines represents the time when the superradiance ends. We set $G_N=1$, $c=0.1$.}
\label{page curve}
\end{figure}

\qquad So far, we solve the information paradox of non-extremal BTZ black holes and obtain the Page time \eqref{page time} and the scrambling time \eqref{scrambling time}. At first sight, it seems that they are both divergent at the extremal case $(J=M)$. However, there is a fundamental difference between extremal black holes and non-extremal black holes due to the different causal structure \cite{extremal bh}. One can not simply understand that the Page time is divergent in the extremal case from the continuous limit of the non-extremal case. Some works reporting on extremal black holes \cite{extremal,charged dilaton1}, which show that both the entanglement entropy and the Page time are ill-defined at the zero temperature.\\
\indent Therefore, in order to avoid this problem, we consider the superradiance of the system. Superradiance can extract rotational energy from black holes, which is analog of the famous Penrose process. In this section, we first prove that the system has the superradiance phenomenon. Later, we analyze its effect on the previous results through numerical calculations.\\
\indent We mainly consider the scalar fields here\footnote{For the fermionic fields, the Pauli exclusion principle prevents fermions be amplified. See \cite{fermi}}. The scalar field $\varphi$ satisfies the Klein-Gorden equation
\begin{equation}
\big( \square - \lambda R-\mu^2 \big)\varphi=0, \label{KG eq}
\end{equation}
where $\lambda$ is the conformal coupling constant, which is vanishing when the scalar field is minimally coupled with the background metric. The curvature is given by $R=-\frac{6}{\ell^2}$ and $\mu$ can be regarded as the mass of field. $\square$ represents the D'Almbert operator, which is given by
\begin{equation}
\square \equiv \frac{1}{\sqrt{-g}} \partial_{\mu} \big( \sqrt{-g}g^{\mu \nu} \partial_{\nu} \big). \label{operator}
\end{equation}
Then the equation \eqref{KG eq} can be rewritten as
\begin{equation}
\frac{1}{\sqrt{-g}} \partial_{\mu} \big( \sqrt{-g}g^{\mu \nu} \partial_{\nu} \varphi \big) - \mu^{\prime} \varphi =0, \label{btz KG}
\end{equation}
where $\mu^{\prime}=\mu^2+\lambda R$. Make the ansatz
\begin{equation}
\varphi (t,r, \phi)=e^{i (\omega t+m \phi)} \frac{R(r)}{\sqrt{r}}, \label{ansatz}
\end{equation}
where $\omega$ is the frequency or energy and $m$ is the axial quantum number or angular momentum. The radial wavefunction $R(r)$ satisfies the following equation
\begin{equation}
\frac{d^2R(r)}{dr^{\ast 2}} +V_{\text{eff}} R(r)=0, \label{radial eq}
\end{equation}
where $r^{\ast}$ is the tortoise coordinate \eqref{tortoise}, and the effective potential $V_{\text{eff}}$ can be written as \cite{BTZ superradiance}
\begin{equation}
V_{\text{eff}}= \bigg \{ \bigg (\omega +m N^{\phi}\bigg )^2-f(r) \bigg[ \frac{m^2}{r^2}+\mu^{\prime 2}+ \frac{1}{2} \frac{d}{dr} \bigg[ \frac{f(r)}{r^{\frac{3}{2}}} \bigg] + \frac{f(r)}{2r^{\frac{2}{5}}} \bigg] \bigg \}.
\end{equation}
Here $f(r)$ and $N^{\phi}$ are metric functions \eqref{metric function} and \eqref{angular function}.\\
\indent Let us analyze the asymptotic solution of the radial equation \eqref{radial eq} at the event horizon and the infinity. At the event horizon $r=r_+$, the metric function $f(r_+)=0$, so the radial equation \eqref{radial eq} reduces as
\begin{equation}
\frac{d^2R(r_+)}{dr^{\ast 2}(r_+)} + \omega_1^2 R(r_+)=0,  \label{horizon R}
\end{equation}
with
\begin{equation}
\omega_1=\omega+mN^{\phi}(r_+) =\omega-m\Omega_H,   \label{w1}
\end{equation}
where $\Omega_H$ is the velocity of angular at the event horizon \eqref{angular velocity}. Therefore, the radial wavefunction $R(r_+)$ is obtained at the event horizon
\begin{equation}
R(r_+) \sim e^{\pm i \omega_1 r^{\ast}}.    \label{R1}
\end{equation}
Because we couple baths and make the transparent boundary, the wave can spread to the future null infinity. At the infinity $r \to \infty$, the metric function $f(\infty)=1$ is the Minkowski spacetime. Therefore, the behavior of $R(\infty)$ is given by the similar analysis
\begin{equation}
R(\infty) \sim e^{\pm i \omega_2 r^{\ast}}, \label{R2}
\end{equation}
with the frequency
\begin{equation}
\omega_2=\sqrt{\big(\omega + mN^{\phi}(\infty)\big)^2-\mu^{\prime 2}}= \sqrt{\omega^2-\mu^{\prime 2}}.  \label{w2}
\end{equation}
For the radial wavefunction $R(r)$, we have
\begin{equation}
 \left \{
\begin{array}{lr}
R_1(r^{\ast})=e^{-i \omega_2 r^{\ast}} + \mathcal{R}e^{i \omega_2 r^{\ast}},    \hspace{2.0cm}  r \to \infty &\\
R_2(r^{\ast})=\mathcal{T} e^{-i \omega_1 r^{\ast}},  \hspace{3.5cm} r \to r_+&
\end{array}
\right. \label{superradiance analysis}
\end{equation}
where $\mathcal{R}$ and $\mathcal{T}$ are the reflection and the transmission coefficients, respectively. One can easily obtain the following expression from \eqref{radial eq}
\begin{equation}
W=R_1 \frac{dR_2}{dr^{\ast}}-R_2\frac{dR_1}{dr^{\ast}}, \label{condition}
\end{equation}
where $W$ is the Wronskian. On the other hand, the complex conjugate $R_{1,2}^{\ast}$ is another linearly independent solution. Evaluating the above equation, we find
\begin{equation}
|\mathcal{R}|^2=1-\frac{\omega_1}{\omega_2}|\mathcal{T}|^2.  \label{superradiance eq}
\end{equation}
Therefore, for $\omega_1<0$ or
\begin{equation}
0<\omega<m\Omega_H, \label{superradiance region}
\end{equation}
then the reflection coefficient $|\mathcal{R}|^2>1$. In other words, the reflected wave has a bigger amplitude than the incident wave as the superradiance is continues. The essence of the amplification effect of the incident wave is a kind of stimulated radiation, which is the non-thermal radiation that is not related to the temperature \eqref{temperature} of black holes. At the same time, the superradiance will cause the angular momentum of black holes to decrease so that at the end, a rotating black hole will become a static black hole.\\
\indent Now, we study the effect of the superradiance on the previous results by numerical calculations. The rate of the angular momentum loss is given by the following formula \cite{angular loss}
\begin{equation}
\frac{dJ}{dt}=-\frac{1}{2\pi} \sum_{l,m} \int_{0}^{m \Omega_H} m (|\mathcal{R}|^2-1) d\omega,  \label{angular momentum loss}
\end{equation}
where the $l$ is the spherical harmonic. On the other hand, the thermal Hawking radiation and non-thermal superradiance exist at the same time in the system. We assume that baths only replenishes energy (mass) to the black hole and is always in thermal equilibrium with the black hole. In this way, we can obtain the loss rate of the mass $\frac{dM}{dt}$ through
\begin{equation}
\frac{dT}{dt}=\frac{dT}{dJ} \frac{dJ}{dt} +\frac{dT}{dM} \frac{dM}{dt} \equiv 0.  \label{mass loss}
\end{equation}
We give the numerical results only for $l=m=1$ mode which is the most dominant and set $\mu^{\prime 2}=0$ for convenience.\\
\indent Setting the initial conditions $M(0)=1$ and different $J(0)$. The angular momentum, the mass and the black hole entropy changes are shown in \mpref{bh parameters}.
\\
The greater the angular momentum, the more dominant the superradiance (\mpref{angualr}) and the faster the mass and entropy decrease (\mpref{mass} and \mpref{BH entropy}). The greater the initial angular momentum, the shorter the superradiance time. This is similar to the result given in \cite{particle emission}. The effect of superradiance on the Page time \eqref{page time} and the scrambling time \eqref{scrambling time} is plotted in \mpref{tp and tscr}.
\\
We find that the superradiance causes both the Page time and the scrambling time to decrease. The greater the angular momentum, the faster they decrease. Interestingly, when the superradiance ends, the static BTZ black hole is restored, its Page time $(t_{\text{Page}}(\text{static}) \equiv \frac{3 \pi}{2cG_N})$ does not depend on the mass and is a constant (\mpref{tpage}). This is consistent with \cite{btz}.\\
\indent At last, we give the Page curve after considering the superradiance, which is shown in \mpref{page curve}. When the central charge is not very large $c \ll 1/G_N$, the period of superradiance is much shorter than the Page time. After the superradiance ends, the Hawking radiation dominates. Because the black hole is in thermal equilibrium with the bath, the slope of each curve before the Page time $t_1$ is the same as the case without considering the superradiance (black lines). Moreover, the greater the initial angular momentum, the shorter the period of superradiance. Therefore, a rotating BTZ black hole quickly becomes a static BTZ black hole, which has the smallest Page time. Thus the ill-definedness problem in the extremal case does not occur. In fact, the Hawking radiation becomes purely superradiant in the extremal limit \cite{particle emission,angular loss}.

\section{Discussion and Conclusion} \label{Conclusion}
\qquad In summary, we study the black hole information paradox for BTZ black holes. Considering the reason that the BTZ spacetime is local AdS$_3$, in order to make it radiative, we adopt the technique of coupling auxiliary baths. So the behavior of the entanglement entropy of Hawking radiation can be studied by the island formula \eqref{island formula}. For the non-extremal case, there is no island at early times, and the entanglement entropy is proportional to time and exceeds the entropy bound \eqref{result1}. However, the island outside the event horizon \eqref{island location} leads to the entropy eventually reaching a saturating value at late times \eqref{result2}. For these processes, the entropy obeys the Page curve. In addition, we pay attention to the effect of angular momentum on these results. Related works \cite{extremal,charged dilaton1} report the situation of extremal black holes and show that the boundary of the island hits the curvature singularity where the semiclassical approximation is invalid. Thus, both the entanglement entropy and the Page time are ill-defined in the extremal case. However, it does not seem to be applicable to the BTZ black hole spacetime, because BTZ black holes has no curvature singularity at the origin \cite{btz review}. Therefore, we consider the superradiance that can extract rotating energy from BTZ black holes and makes black hole less extremal, the results show that a rotating BTZ black hole eventually becomes a statically neutral BTZ black hole. In this process, both the Page time and the scrambling time decrease and become finite. The Page curve after superradiant correction is shown in \mpref{page curve}.\\
\indent In the future study, we will pay attention to the following points:\\
(1) The aspect of the background. The object of this paper is the three-dimensional BTZ spacetime. On the one hand,  it is different from the asymptotically flat spacetime where there is no need to couple the auxiliary bath to make black holes radiative. On the other hand, we do not consider the case of coupled the Maxwell field. Therefore, we expect the analysis in this paper could be extended to the most general four-dimensional Kerr-Newman spacetime. This will also be very interesting.\\
(2) The aspect of the construction. This paper only considers the construction of a single island. Although this is sufficient to solve the paradox, the construction with multiple islands can also be considered. Multiple islands imply multiple extremal surfaces, which may cause more phase transitions of the entanglement entropy and describes the Page curve in more detail.\\
(3) The aspect of the quantum information. There exist a traversable wormhole between the island and the radiation region in the spirit of ``ER=EPR" \cite{EPR}. There is a protocol called quantum teleportation \cite{wormholes} to extract the information residing on the island. We can create the negative energy shock wave in the bulk and the information transfers to the bath \cite{eternal bh}. However, the large amount of shock wave with negative energy not only backreacts on the bulk geometry but also affects the
(averaged) null energy condition. Thus, this operation can not be continued.\\
\indent In all, we still need to study the island further, which could reveal the nature of quantum gravity.

\section*{Acknowledgement}
We would like to thank Hai-Ming Yuan, Yu-Qi Lei and Qing-Bin Wang for helpful discussions. The study was partially supported by NSFC, China (grant No.11875184). The work of Sang-Jin Sin is partially supported by Mid-career Researcher Program through the National Research Foundation of Korea grant No. NRF2021R1A2B5B0200260.

\end{document}